\input{aipcheck}

\documentclass[
    ,final            % use final for the camera ready runs
  ]
  {aipproc}

\layoutstyle{6x9}
\usepackage{rotate}
\usepackage{amsmath}
\usepackage{amsfonts}
\usepackage{graphicx}
\usepackage{graphics}
\usepackage{epsfig}
\begin{document}

\title{BARRED GALAXIES: STUDYING THE CHAOTIC AND ORDERED NATURE OF ORBITS }


\author{T. Manos}{address={Center for Research and Applications of Nonlinear Systems
(CRANS), Department of Mathematics, University of Patras, GREECE.},
altaddress={Observatoire Astronomique de Marseille-Provence (OAMP),
2 Place Le Verrier, 13248 Marseille cedex 04, FRANCE.} }

\author{E. Athanassoula}{address={Observatoire Astronomique de
    Marseille-Provence (OAMP), 2 Place Le Verrier, 13248 Marseille
    cedex 04, FRANCE.}}

\begin{abstract}
The chaotic or ordered character of orbits in galactic models is an
important issue, since it can influence dynamical evolution. This
distinction can be achieved with the help of the \emph{\textbf{S}}maller
\emph{\textbf{AL}}ingment \emph{\textbf{I}}ndex - (\emph{\textbf{SALI}}).
We describe here briefly this method and its advantages. Then we apply
it to that case of 2D and 3D barred galaxy potentials. In particular,
we find the fraction of \textbf{chaotic and ordered orbits} in such potentials
and present how this fraction changes when the main parameters of the
model are varied. For this, we consider models with different bar mass, bar
thickness or pattern speed. Varying only one parameter at a time, we
find that bars that are more massive, or thinner, or faster, have a
larger fraction of chaotic orbits.

%The chaotic or ordered character of the orbits in galactic models is
%a very important issue in Astronomy. This disjunction can be done by
%using the \emph{\textbf{S}}maller \emph{\textbf{AL}}ingment
%\emph{\textbf{I}}ndex - (\emph{\textbf{SALI}}), which is a very
%useful tool for the qualitative separation of the \textbf{chaotic or
%ordered motion} inside them. Here, we use the SALI in 2D and 3D
%\textbf{barred galactic potentials}, which is defined by the
%Ferrers' potential and we study the behavior of the model while some
%substantial parameters vary. In order to achieve this goal, we
%calculate the fraction of the chaotic and ordered trajectories of
%the model, as the mass of the various components, the bar axial
%ratio and the pattern speed of the bar component changes.
\end{abstract}

\maketitle

%%%%%%%%%%%%%%%%%%%%%%%%%%%%%%%%%%%%%%%%%%%%
%% MAINMATTER
%%%%%%%%%%%%%%%%%%%%%%%%%%%%%%%%%%%%%%%%%%%%

\section{INTRODUCTION}
%The distinction between ordered and chaotic motion in dynamical
%systems is fundamental in many areas of applied sciences. This
%distinction is particularly difficult in systems with many degrees
%of freedom (dof), basically because it is not feasible to visualize
%their phase space. So, we need fast and accurate tools to give us
%in-formation about the chaotic or ordered character of orbits,
%especially for conservative systems.

The chaotic and ordered motion in dynamical systems is a very
important issue in non - linear sciences in general and in dynamical
astronomy in particular. The qualitative distinction of the motion is, generally,
not that trivial and it becomes harder in systems of many degrees of
freedom.  Thus, it is necessary to use fast and precise methods
to indicate the nature of the orbits.

Many methods have been developed over the years trying to give an
answer to this problem. The inspection of the successive
intersections of an orbit with a Poincar\'{e} surface of section
(PSS) \cite{LL} has been used mainly for 2 dimensional (2D) maps and
2 degrees of freedom Hamiltonian systems. One of the most common methods of chaos
detection is the computation of the "Maximal Lyapunov Characteristic
Number" (LCN) \cite{Fro:1},\cite{Ben:1}, which can be applied for
systems with many degrees of freedom. Another very efficient method
is the "Frequency Map Analysis"
\cite{Las:1},\cite{Las:2}. In recent years new methods have
been introduced such as the study of spectra of "Short Time Lyapunov
Characteristic Numbers" \cite{Fro:2},\cite{Loh:1} or "Stretching
Numbers" \cite{Vog:1},\cite{Con:1} and the "Spectral Distance" of
such spectra \cite{Vog:2}, as well as the study of spectra of
helicity and twist angles \cite{Con:2},\cite{Con:3},\cite{Fro:3}. In
addition, Froeschl\'{e}, Lega \& Gonzi \cite{Fro:4} introduced the "Fast
Lyapunov Indicator" (FLI), while Vozikis, Varvoglis \& Tsiganis
\cite{Voz:1} proposed
a method based on the frequency analysis of "Stretching Numbers".

Recently, a new, fast and easy to compute indicator of the chaotic
or ordered nature of orbits, has been introduced \cite{sk:1} and was
applied successfully in different dynamical
systems \cite{sk:1},\cite{sk:2},\cite{sk:3},\cite{sk:4},\cite{sk:5},\cite{Pan:1},\cite{Ant:1},\cite{SESS},
the so-called "\textbf{S}maller \textbf{Al}ignment \textbf{I}ndex
(\textbf{SALI})", or, as elsewhere called \textbf{A}lignment
\textbf{I}ndex \textbf{(AI)} \cite{VKS},\cite{KVC}. In the present
paper, we first recall its definition
and then we show its effectiveness in distinguishing between ordered
and chaotic motion, by applying it to a barred potential of 2 and 3
degrees of freedom.

\section{DEFINITION OF THE SMALLER ALIGNMENT INDEX}
Let us consider the  $n$ - dimensional phase space of a conservative
dynamical system, which could be a symplectic map or a Hamiltonian
flow. We consider also an orbit in that space with initial condition $P(0)=(x_{1}(0),x_{2}(0),...,x_{n}(0))$
and a deviation vector $v(0)=(dx_{1}(0),dx_{2}(0),...,dx_{n}(0))$ from the initial point $P(0)$. In order to compute
the SALI for a given orbit one has to follow the time evolution of
the orbit with initial condition itself and also of two deviation
vectors $v_{1}(t),v_{2}(t)$ which initially point in two different directions. The
evolution of these deviation vectors is given by the variational
equations for a flow and by the tangent map for a discrete-time
system. At every time step the two deviation vectors $v_{1}(t)$ and $v_{2}(t)$ are
normalized and the SALI is then computed as:

 \begin{equation}\label{eq:SALI}
    SALI(t)=min \left\{\left\|\frac{v_{1}(t)}{\|v_{1}(t)\|}+\frac{v_{2}(t)}{\|v_{2}(t)\|}\right\|,\left\|\frac{v_{1}(t)}{\|v_{1}(t)\|}-\frac{v_{2}(t)}{\|v_{2}(t)\|}\right\|\right\}.
\end{equation}

The properties of the time evolution of the SALI clearly distinguish
between ordered and chaotic motion as follows: In the case of
Hamiltonian flows or  dimensional symplectic maps with $n \geq 2$, the SALI
fluctuates around a non-zero value for ordered orbits, while it
tends to zero for chaotic
orbits \cite{sk:1},\cite{sk:3},\cite{sk:4}. In the case
of 2D maps the SALI tends to zero both for ordered and chaotic
orbits, following however completely different time
rates, which again allows us to distinguish between the
two cases. In 2 and 3 degrees of freedom Hamiltonian systems the distinction
between ordered and chaotic motion is easy because the ordered
motion occurs on a 2D or 4D torus respectively on which any initial
deviation vector becomes almost tangent after a short transient
period. In general, two different initial deviation vectors become
tangent to different directions on the torus, producing different
sequences of vectors, so that SALI does not tend to zero but
fluctuates around positive values. On the other hand, for
chaotic orbits, any two initially different deviation vectors tend
to coincide in the direction defined by the nearby unstable manifold
and hence either coincides with each other, or become opposite. This
means that the SALI tends to zero when the orbit is chaotic and to a
non-zero value when the orbit is ordered. Thus, the completely
different behavior of the SALI helps us to distinguish between
ordered and chaotic motion in Hamiltonian systems with 2 and 3 degrees of freedom
and in general in dynamical systems of higher dimensionality.

\section{THE MODEL}
A 3D rotating bar model can be described by the Hamiltonian
function:

\begin{equation}\label{eq:Hamilton}
   H=\frac{1}{2} (p_{x}^{2}+p_{y}^{2}+p_{z}^{2})+ V(x,y,z) -
   \Omega_{b} (xp_{y}-ypx_{x}),
\end{equation}
where $x$ is the long axis, $y$ the intermediate and $z$ the short
axis around which the bar rotates. The $p_{x},p_{y}$ and $p_{z}$ are the
canonically conjugate momenta. Finally, $V$ is the potential,
$\Omega_{b}$ represents the pattern speed of the bar and $H$ is the
total energy of the system.

The potential of the model $V$ consists of three components:

a) A \textit{disc}, represented by a Miyamoto disc \cite{Miy.1975}:
\begin{equation}\label{Miy_disc}
  V_D=- \frac{G M_{D}}{\sqrt{x^{2}+y^{2}+(A+\sqrt{z^{2}+B^{2}})^{2}}}.
\end{equation}
where $M_{D}$ is the total mass of the disc, $A$ and $B$ are the
horizontal and vertical scalelengths, and $G$ is the gravitational
constant.

b) The \textit{bulge} is modeled by a Plummer sphere with the
potential:
\begin{equation}\label{Plum_sphere}
    V_S=-\frac{G M_{S}}{\sqrt{x^{2}+y^{2}+z^{2}+\epsilon_{s}^{2}}}
\end{equation}
where $\epsilon_{s}$ is the scalelength of the bulge and $M_{S}$ is
its total mass.

c) The triaxial Ferrers \textit{bar}, the density $\rho(x)$ of which
is:
\begin{eqnarray}\label{Ferr_bar}
  \rho(x)=\begin{cases}\rho_{c}(1-m^{2})^{2} &, m<1  \\
              \qquad 0 &, m\geq1 \end{cases},
\end{eqnarray}

\noindent
where $\rho_{c}=\frac{105}{32\pi}\frac{G
M_{B}}{abc}$ the central density, $M_{B}$ is the total mass of
the bar and

\begin{equation}\label{Ferr_m}
  m^{2}=\frac{x^{2}}{a^{2}}+\frac{y^{2}}{b^{2}}+\frac{z^{2}}{c^{2}},
\qquad a>b>c\geq 0,
\end{equation}

with $a,b$ and $c$ are the semi-axes and $M_{B}$ the mass of the bar
component. The corresponding potential $V_{B}$ and the forces are
given in D. Pfenniger (1984) \cite{Pfe}.

\subsection{Applications in the 2D case}
We first apply the SALI index to the 2 degrees of freedom barred potential. In this
case, we can have a Poincar\'{e} Surface of Section (PSS) and can
check the effectiveness of the SALI, by comparing the results.

In figure 1, we present two orbits of different kind: (i)
$(x,y,p_{x},p_{y})=(1.5,0,p_{x}(H),0)$ and (ii)
$(x,y,p_{x},p_{y})=(-0.9,0,p_{x}(H),0)$. For this application
$H=-0.3$. In panels a) and b) of figure 1, we show the orbit
projections in the $(x,y)$-plane and in panel c) we draw their
corresponding PSS, where we can see that the chaotic orbit (i) tends
to fill with scattered points the available part of the
$(x,\dot{x})$ plane and that the ordered orbit (ii) creates a closed
invariant curve. Finally, in panel d) we apply the SALI method for
these two orbits. For the chaotic one, the SALI tends to zero
$(10^{-16})$ exponentially after some time steps while for the
regular orbit, it fluctuates around a positive number. By choosing
initial conditions on the line $\dot{x}=0$ of the PSS and
calculating the values of the SALI, we can detect very small regions
of stability that can not be visualized easily by the PSS method.
Repeating this for many values of the energy, we are able to follow
the change of the fraction of chaotic and ordered orbits in the
phase space as the energy of the orbit varies.

%\clearpage
\begin{figure}
  \includegraphics[height=0.22\textheight]{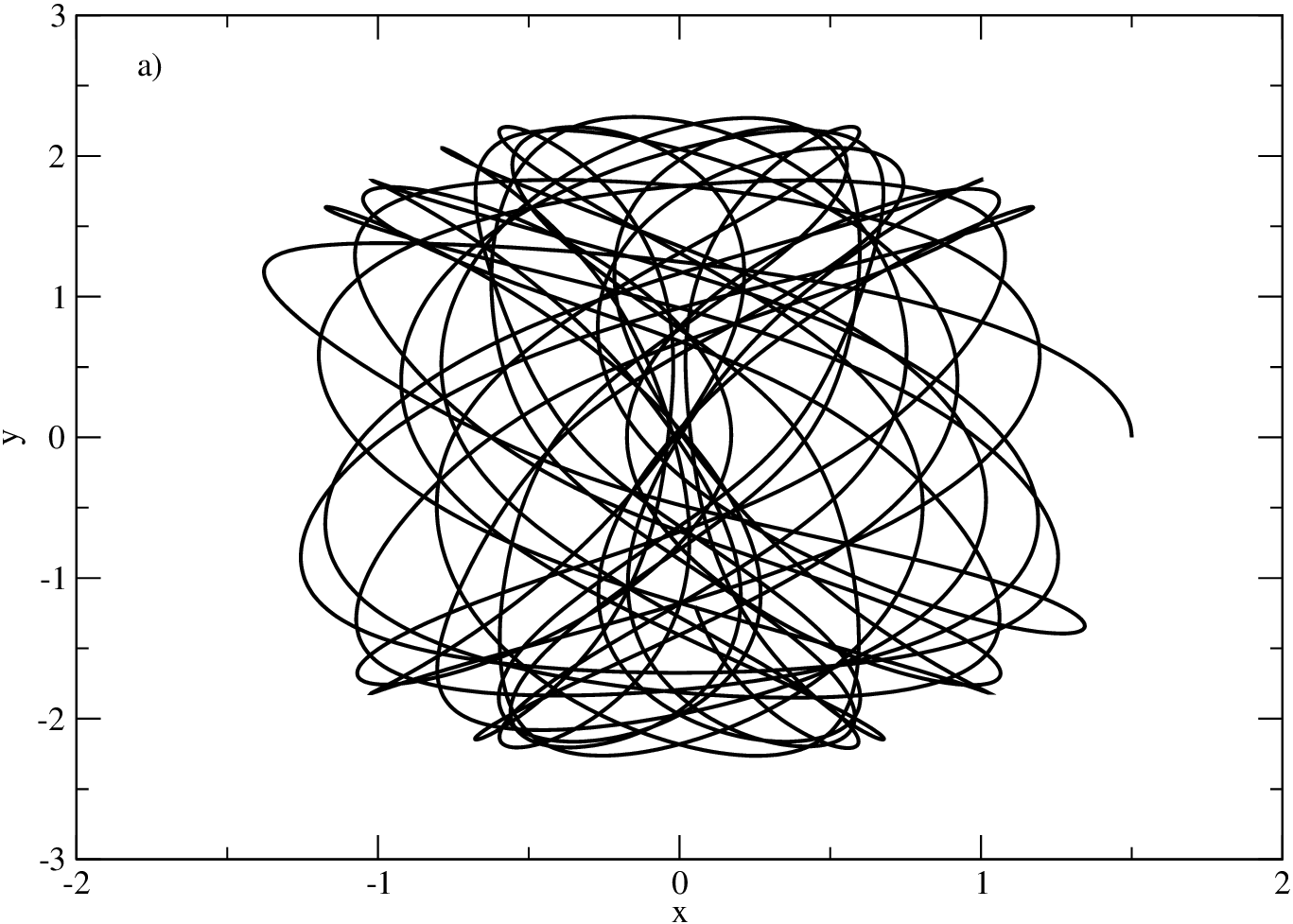}\hspace{0.5cm}
  \includegraphics[height=0.22\textheight]{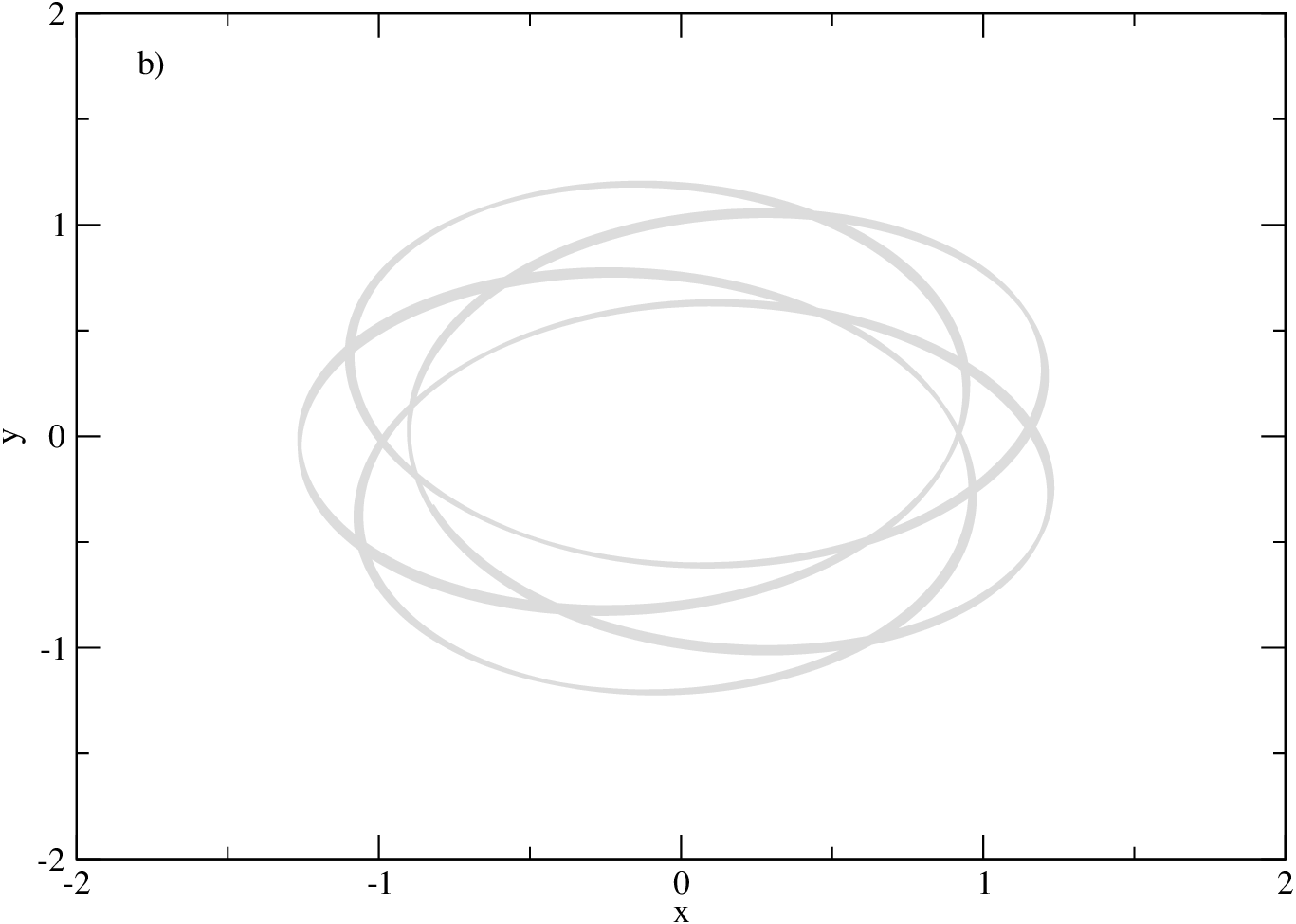}
\end{figure}
\begin{figure}
  \includegraphics[height=0.21\textheight]{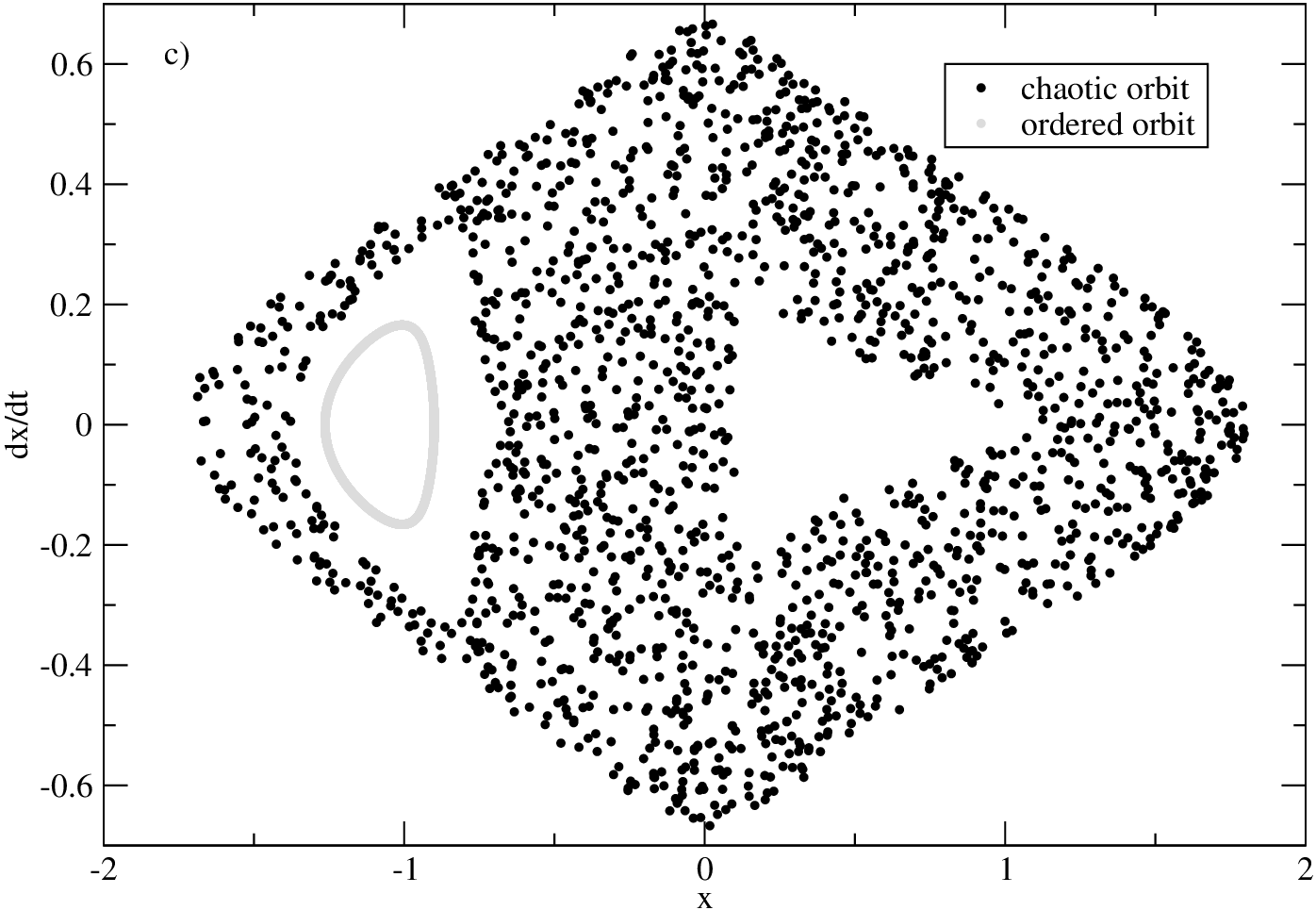}\hspace{0.5cm}
  \includegraphics[height=0.21\textheight]{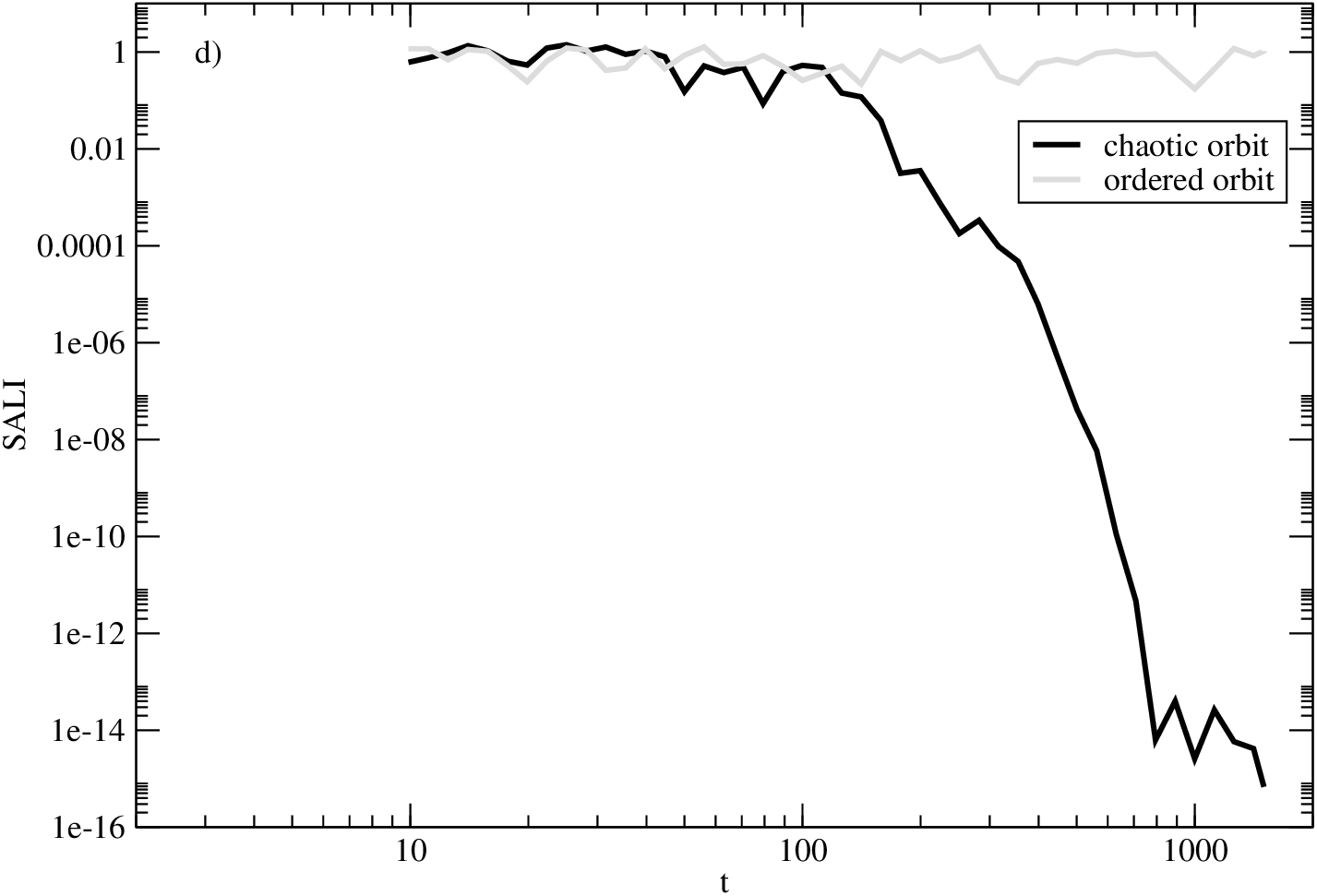}
  \caption{Projections of the orbits (i) and (ii) in
the $(x,y)$-plane (panels a), b)). PSS of these two orbits in panel
c). Behaviour of the SALI of the same orbits (panel d)).}
\end{figure}

%\clearpage
\subsection{Applications in the 3D case}
In figure 2 we present percentages of chaotic, intermediate and
regular trajectories, where we vary the mass of the bar component
(panels A3-B3) and the length of the short $z$-axis (panels A2-B2)
of the initial models A1 and B1. The two rows differ in the way we
give the 27000 initial conditions. For the A1, we give initial
conditions in the plane $(x,p_{y},z)$ with $(y,p_{x},p_{z})=(0,0,0)$
and for the B1, in the plane $(x,p_{y},p_{z})$ with
$(y,z,p_{x})=(0,0,0)$. By comparing the results, we see that the
increase of the bar mass causes more chaotic behavior in both cases
(panels A3, B3). This confirms the results by Athanassoula et al.
(1983) \cite{ABMP} in 2 degrees of freedom. On the other hand, it is
obvious

%%%%%%%%%%%%%%%%%%%%%%%%%%%%%%%%%%%%%%%%%%%%
%% Sample figure:
%%
%% The option [height=...] scales the picture to the given height,
%% without it it would be printed at its nominal size
%%%%%%%%%%%%%%%%%%%%%%%%%%%%%%%%%%%%%%%%%%%%
\clearpage
\begin{figure}
  \includegraphics[height=0.25\textheight]{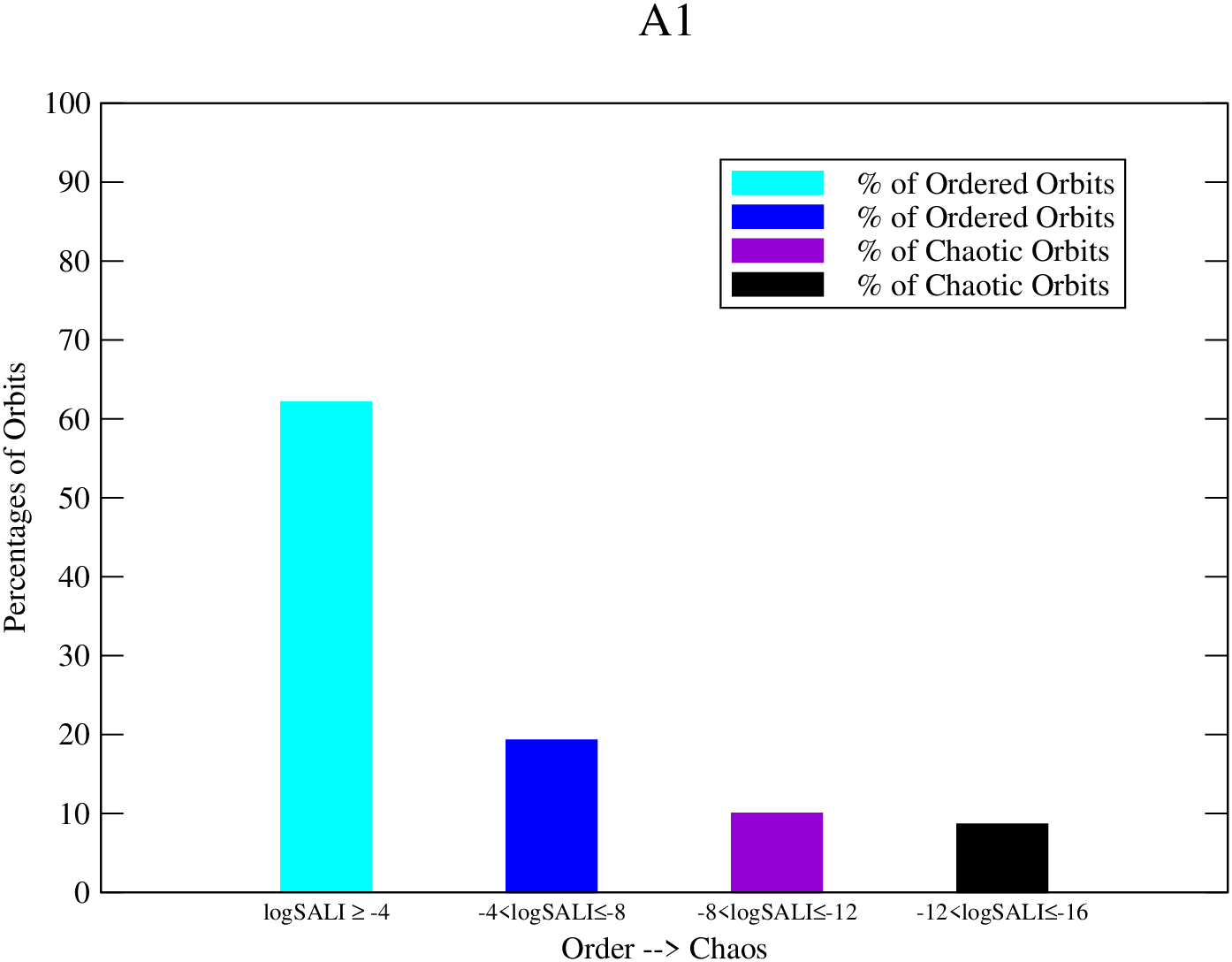}\hspace{0.3cm}
  \includegraphics[height=0.25\textheight]{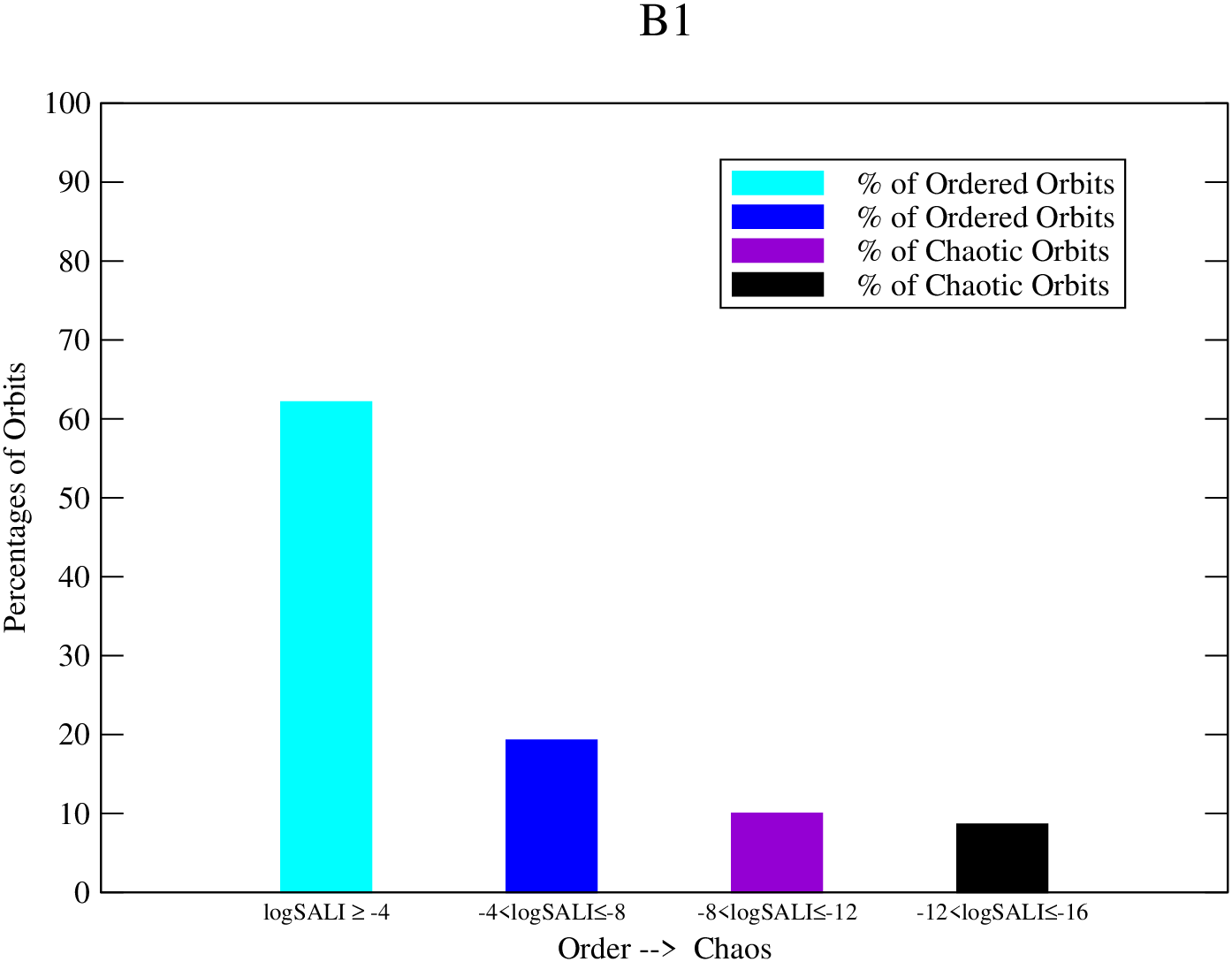}\hspace{0.3cm}
%  \caption{Picture to fixed height}
\end{figure}
\begin{figure}
\includegraphics[height=0.25\textheight]{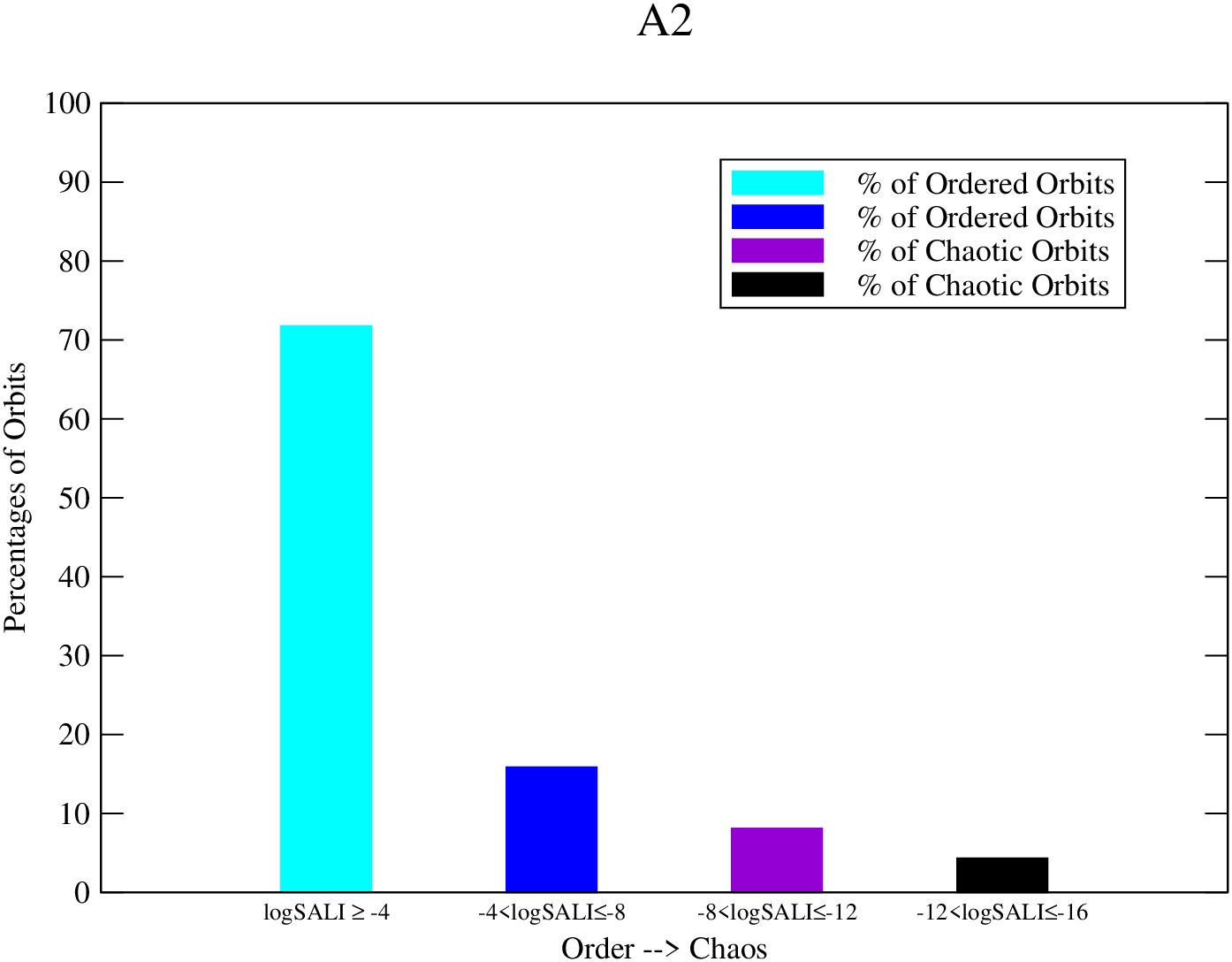}\hspace{0.3cm}
  \includegraphics[height=0.25\textheight]{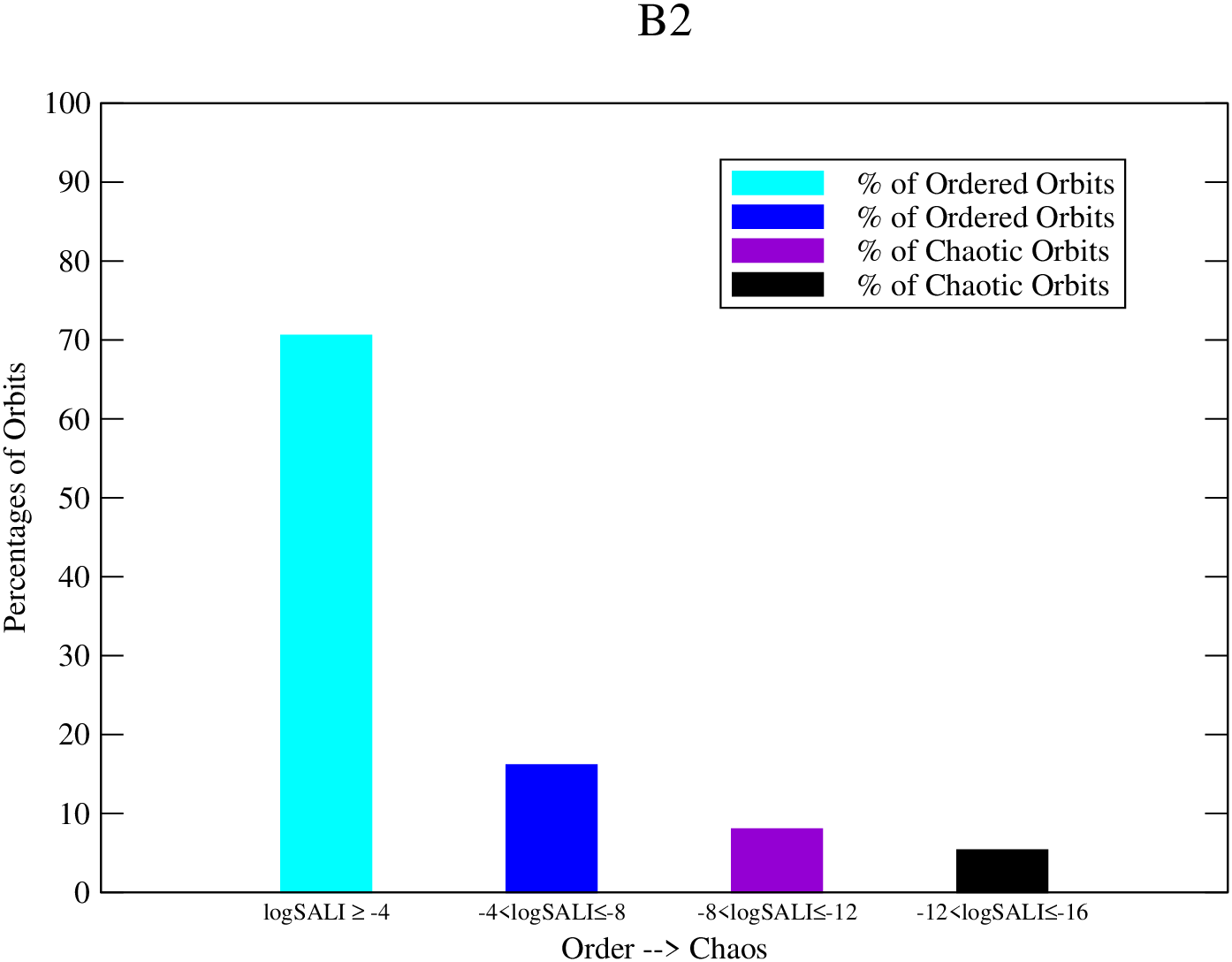}\hspace{0.3cm}
\end{figure}
\begin{figure}
\includegraphics[height=0.25\textheight]{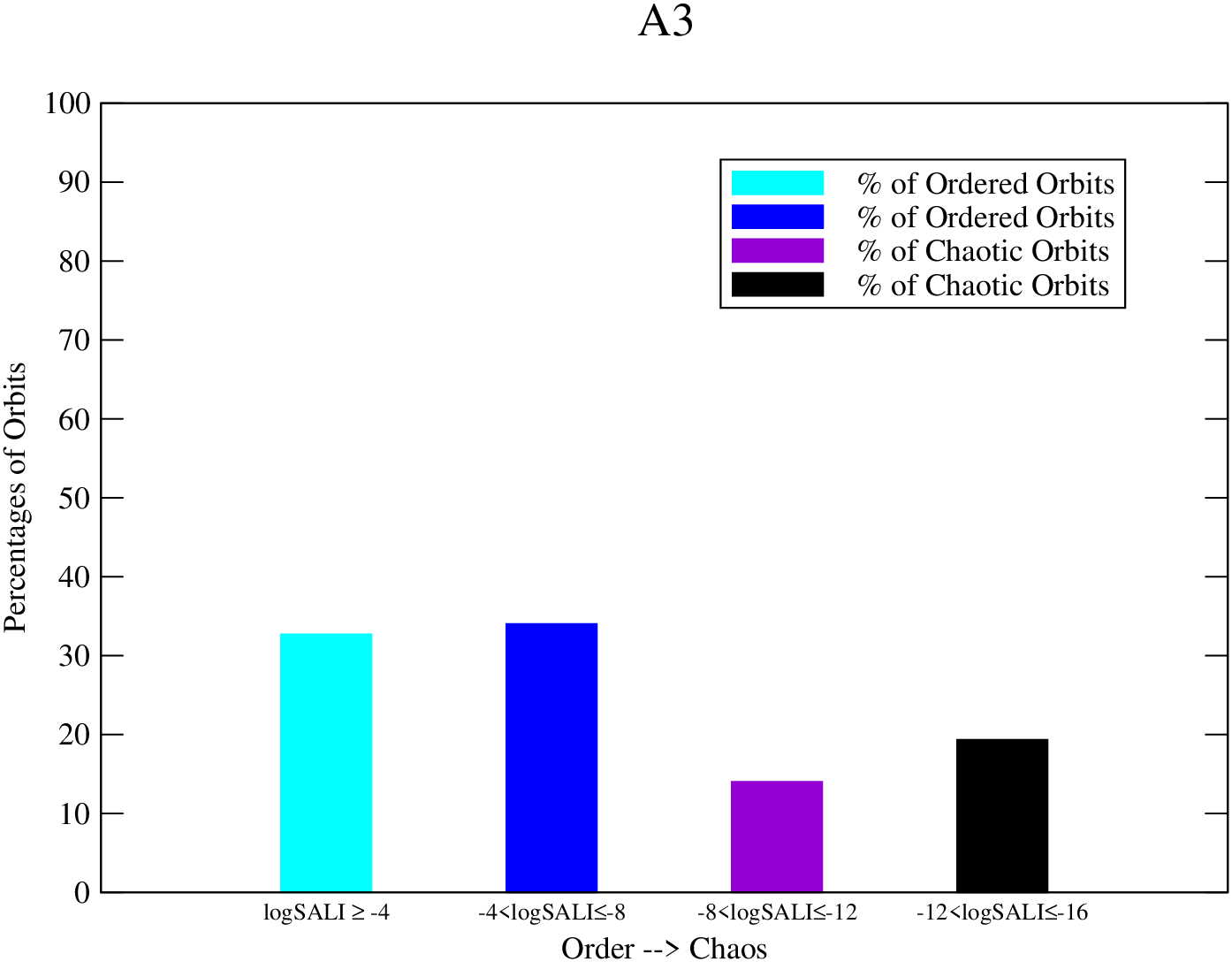}\hspace{0.3cm}
  \includegraphics[height=0.25\textheight]{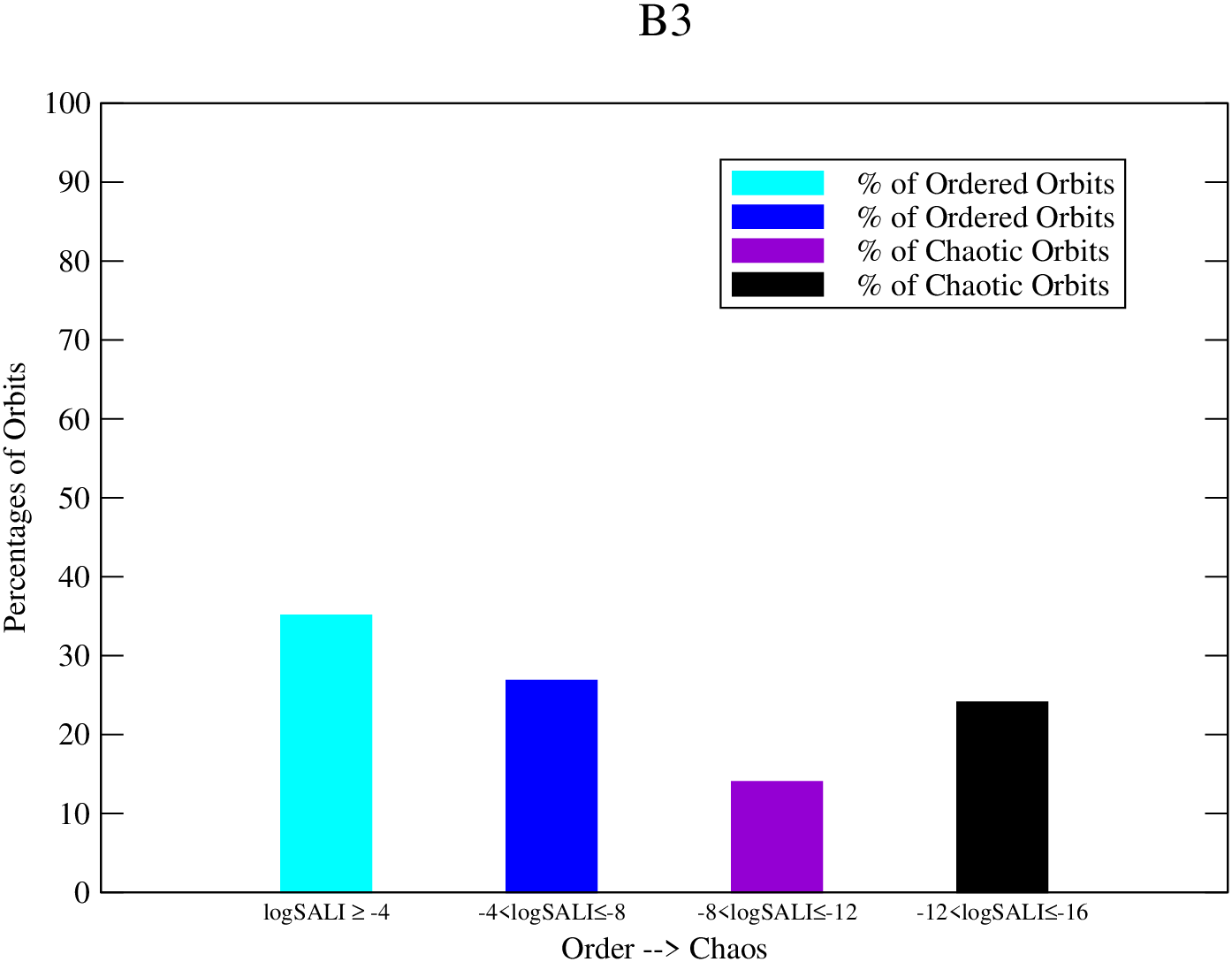}\hspace{0.3cm}
      \caption{Percentages of regular (first and second bar) and chaotic (third and forth bar) orbits,
      for our standard model (first row of panels), a model with a thick bar (second row) and a
      model with a massive bar (third row). The two columns show two different ways of choosing the orbital population.}
   \end{figure}

\clearpage
that when the bar is thicker, i.e. the length of the
$z$-axis larger, the system gets more regular.

Finally, we calculated the same percentages for different pattern
speeds $\Omega_{b}$. From the orientation of periodic orbits,
Contopoulos (1980) \cite{Con:4} showed that bars have to end before
corotation, i.e. that $r_{L}>a$, where $r_{L}$ the Lagrangian, or
corotation, radius. Comparing the shape of the observed dust lanes
along the leading edges of bars to that of the shock loci in
hydrodynamic simulations of gas flow in barred galaxy potentials,
Athanassoula (1992a,b) \cite{Athan:2},\cite{Athan:3} was able to set
both a lower and an upper limit to corotation radius, namely
$r_{L}=(1.2 \pm 0.2)a$. This restricts the range of possible values
of the pattern speed, i.e. $\Omega_{b}=0.0367032$, that corresponds
to the Lagrangian radius $r_{L}=1.4a$ and $\Omega_{b}=0.0554349$,
that corresponds to $r_{L}=1.0a$. Using the extremes of this range,
we investigated how the pattern speed of the bar affects the system
and found that the percentage of the regular orbits is greater in
slow bars.\

%Considering previous orbit calculations
%(Contopoulos, 1980), where it was shown that the value of the
%Lagrangian (corotation) radius is $r_{L}>a$, where $a$ is the length
%of the major axis of the bar, and taking into account hydrodynamic
%simulations (Athanassoula, 1992), it becomes clear that $r_{L}=(1.2
%\pm 0.2)a$. This gives the whole of the available range of the
%$\Omega_{b}$. Having this information, we chose two different values
%for the pattern speed, i.e. $\Omega_{b}=0.0367032$, that correspond
%to the Lagrangian radius $r_{L}=1.4a$ and $\Omega_{b}=0.0554349$,
%that correspond to $r_{L}=1.0a$. It comes out that from the
%percentages of our results that, the high pattern speeds provoke
%more chaotic behavior in the system than lower speeds.

\begin{figure}
  \includegraphics[height=0.22\textheight]{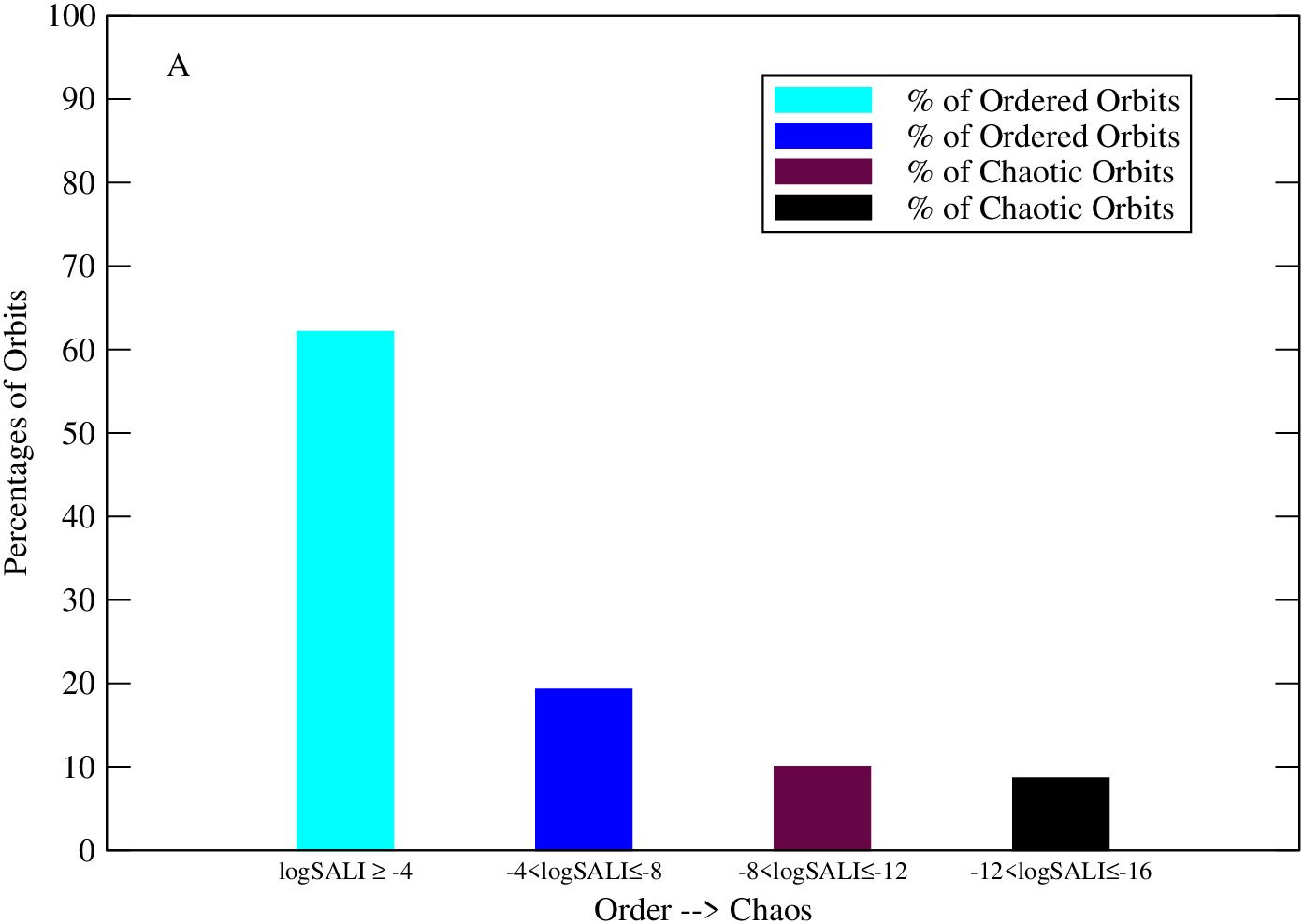}\hspace{0.5cm}
  \includegraphics[height=0.22\textheight]{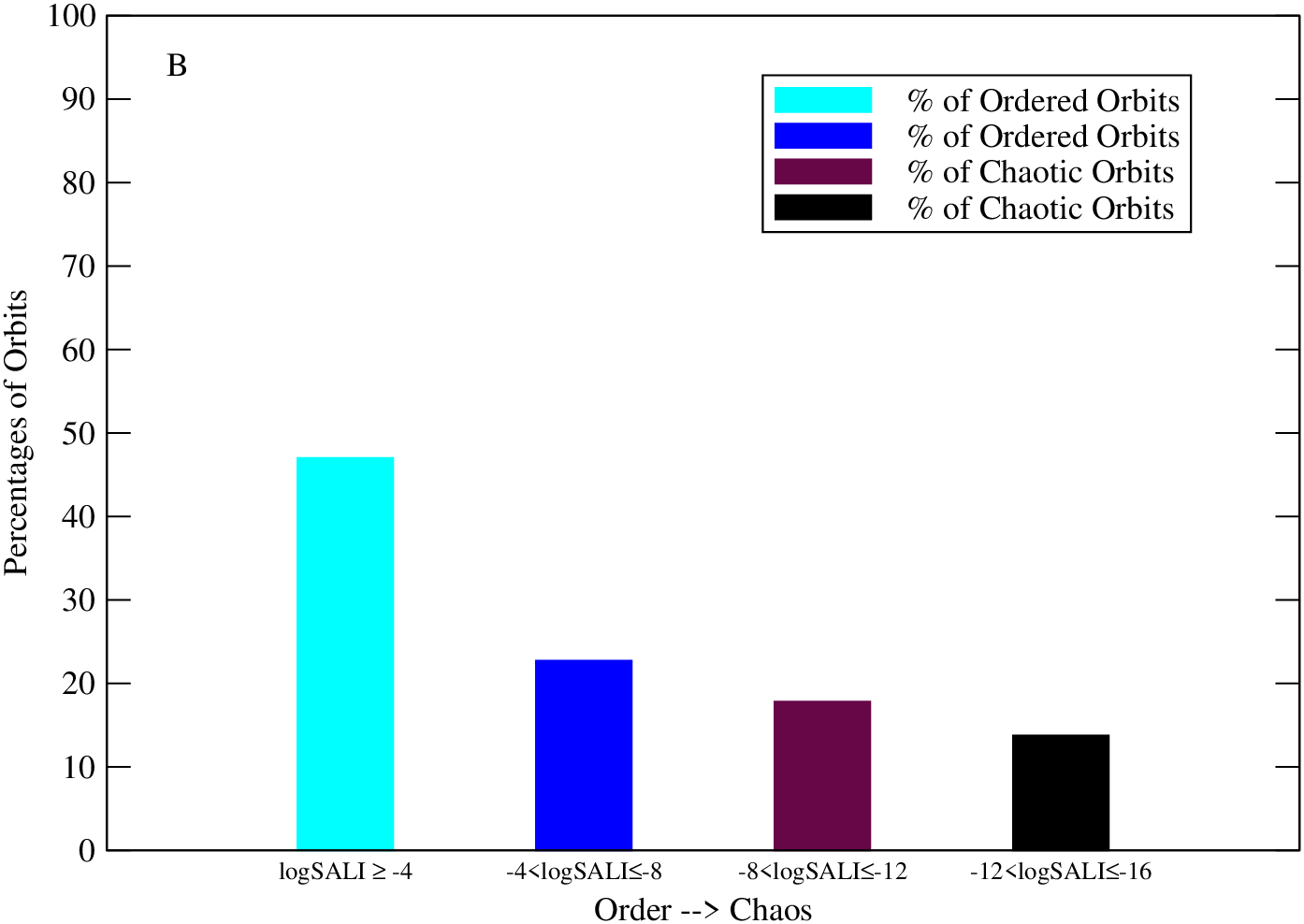}
  \caption{The percentages of the chaotic and ordered orbits for the two different values of the
  pattern speeds. Panel A corresponds to $\Omega_{b}=0.0367032$ and panel B to $\Omega_{b}=0.0554349$. The
  latter presents more chaotic behavior.}
\end{figure}

\section{CONCLUSIONS}
In this paper, we applied the SALI method in the Ferrers barred
galaxy models of 2 and 3 degrees of freedom. We presented and discussed our results
comparing the SALI index with traditional methods, such as the PSS method
for the 2 degrees of freedom and showed its effectiveness. We also, calculated
percentages of chaotic and regular orbits and how they change with
the main model parameters, for the 3 degrees of freedom case. In more detail, we
calculated percentages of chaotic and ordered trajectories, as some
important parameters, as the mass, the length of the short $z$-axis
and the pattern speed of the bar vary in the initial basic model.

%%%%%%%%%%%%%%%%%%%%%%%%%%%%%%%%%%%%%%%%%%%%%%%%
%% BACKMATTER
%%%%%%%%%%%%%%%%%%%%%%%%%%%%%%%%%%%%%%%%%%%%%%%%

\begin{theacknowledgments}
 Thanos Manos was partially supported by "Karatheodory" graduate student fellowship No  B395 of the
 University of Patras and by "Marie - Curie" fellowship No HPMT-CT-2001-00338.
\end{theacknowledgments}

%%%%%%%%%%%%%%%%%%%%%%%%%%%%%%%%%%%%%%%%%%%%%%%%
%% The bibliography can be prepared using the BibTeX program or
%% manually.
%%
%% The code below assumes that BibTeX is used.  If the bibliography is
%% produced without BibTeX comment out the following lines and see the
%% aipguide.pdf for further information.
%%
%% For your convenience a manually coded example is appended
%% after the \end{document}
%%%%%%%%%%%%%%%%%%%%%%%%%%%%%%%%%%%%%%%%%%%%%%%%

%%%%%%%%%%%%%%%%%%%%%%%%%%%%%%%%%%%%%%%%%%%%%%%%
%% You may have to change the BibTeX style below, depending on your
%% setup or preferences.
%%
%%
%% For The AIP proceedings layouts use either
%%%%%%%%%%%%%%%%%%%%%%%%%%%%%%%%%%%%%%%%%%%%

\bibliographystyle{aipproc}   % if natbib is available
%\bibliographystyle{aipprocl} % if natbib is missing

%%%%%%%%%%%%%%%%%%%%%%%%%%%%%%%%%%%%%%%%%%%
%% You probably want to use your own bibtex database here
%%%%%%%%%%%%%%%%%%%%%%%%%%%%%%%%%%%%%%%%%%%
%\bibliography{sample}

\begin{thebibliography}{9}
\bibitem{LL}    M. Lieberman and A. Lichtenberg, \emph{Springer Verlag}, (1992).
\bibitem{Fro:1} C. Froeschl\'{e}, \emph{Cel. Mech.}, \textbf{34}, 95, (1984).
\bibitem{Ben:1} G. Benettin, L. Galgani and J.M. Strelcyn, \emph{Phys. Rev. A}, \textbf{14}, 2338, (1976).
\bibitem{Las:1} J. Laskar, \emph{Icarus}, \textbf{88}, 266, (1990).
\bibitem{Las:2} J. Laskar, C. Froeschl\'{e} and A. Celletti, \emph{Physica D}, \textbf{56}, 253, (1992).
\bibitem{Fro:2} C. Froeschl\'{e}, Ch. Froeschl\'{e} and E. Lohinger, \emph{Celest. Mech. Dyn. Astron.}, \textbf{56}, 307, (1993).
\bibitem{Loh:1} E. Lohinger,  C. Froeschl\'{e} and  R. Dvorak, \emph{Celest. Mech. Dyn. Astron.}, \textbf{56}, 315, (1993).
\bibitem{Vog:1} N. Voglis and G. Contopoulos  \emph{J. Phys. A}, \textbf{ 27}, 4899, (1994).
\bibitem{Con:1} G. Contopoulos, E. Grousousakou and N. Voglis, \emph{Astr. Astroph.}, \textbf{304}, 374, (1995).
\bibitem{Vog:2} N. Voglis, G. Contopoulos and C. Efthymiopoylos, \emph{Celest. Mech. Dyn. Astron.}, \textbf{73}, 211, (1999).
\bibitem{Con:2} G. Contopoulos and  N. Voglis, \emph{Celest. Mech. Dyn. Astron.}, \textbf{64}, 1, (1996).
\bibitem{Con:3} G. Contopoulos and  N. Voglis,  \emph{Astr. Astroph.}, \textbf{317}, 73, (1997).
\bibitem{Fro:3} C. Froeschl\'{e} and E. Lega, \emph{Astr. Astroph.}, \textbf{334}, 355, (1998).
\bibitem{Fro:4} C. Froeschl\'{e}, E. Lega and R. Gonzi, \emph{Celest. Mech. Dyn. Astron}., \textbf{67}, 41, (1997).
\bibitem{Voz:1} Ch. L. Vozikis, H. Varvoglis and K. Tsiganis, \emph{Astr. Astroph.}, \textbf{35}9, 386, (2000).

\bibitem{sk:1}  Ch. Skokos,  \emph{J. Phys. A: Math. Gen.}, \textbf{34}, 10029, (2001).
\bibitem{sk:2}  Ch. Skokos, C. Antonopoulos, T. Bountis  and M. Vrahatis, in \emph{Proceedings of the 4th GRACM}, edited by
                D. T. Tsahalis, \textbf{IV}, 1496. (2002).
\bibitem{sk:3}  Ch. Skokos, C. Antonopoulos, T. Bountis  and M. Vrahatis, \emph{Prog. Theor. Phys. Suppl.}, \textbf{150}, 439, (2003).
\bibitem{sk:4}  Ch. Skokos, C. Antonopoulos, T. Bountis  and M. Vrahatis, in \emph{Proceedings of the Conference Libration Point Orbits and Applications},
                edited by G. Gomez, M. W. Lo and J. J. Masdemont , World Scientific, 653, (2003).
\bibitem{sk:5}  Ch. Skokos, C. Antonopoulos, T. Bountis  and M. Vrahatis,  \emph{J. Phys. A}, \textbf{37}, 6269, (2004).

\bibitem{Pan:1} P. Panagopoulos, T. Bountis and Ch. Skokos, \emph{J. Vib. \& Acoust.}, \textbf{126}, 520, (2004).
\bibitem{Ant:1} C. Antonopoulos, T. Manos  and Ch. Skokos, in \emph{Proceedings of the 1st IC-SCCE}, edited by
                D. T. Tsahalis, Patras Univ. Press, \textbf{III}, 1082, (2005).
\bibitem{SESS}  A. Sz\'{e}ll, B. \'{E}rdi, Zs. S\'{a}ndor and B. Steves, \emph{Mon. Not. R. Astron. Soc.}, \textbf{347}, 380, (2004).


\bibitem{VKS}   N. Voglis , C. Kalapotharakos and I. Stavropoulos, \emph{Mon. Not. R. Astron. Soc.}, \textbf{337}, 619, (2002).
\bibitem{KVC}   C. Kalapotharakos , N. Voglis and  G. Contopoulos, \emph{Mon. Not. R. Astron. Soc.}, \textbf{428}, 905, (2004).

%\bibitem{BGGS}  G. Benettin , L. Galgani , A. Giorgilli and J.-M. Strelcyn, \emph{Meccanica} \textbf{15}, (1980).
%\bibitem{Vog:3}  N. Voglis, G. Contopoulos and C. Efthymiopoylos, \emph{Phys. Rev. E}, \textbf{57}, 372, (1998).

\bibitem{Miy.1975} M. Miyamoto and R. Nagai, \emph{PASJ}, \textbf{27}, 533, (1975).
\bibitem{Pfe}   D. Pfenniger, \emph{Astr. Astroph.}, \textbf{134}, 373, (1984).
\bibitem{ABMP}  E. Athanassoula , O. Bienayme, L. Martinet and D. Pfenniger, \emph{Astr. Astroph}, \textbf{127}, 349, (1983).

\bibitem{Con:4}    G. Contopoulos, \emph{Astr. Astroph}, \textbf{81}, 198, (1980).
\bibitem{Athan:2}  E. Athanassoula, \emph{Mon. Not. R. Astron. Soc.}, \textbf{259}, 328, (1992a).
\bibitem{Athan:3}  E. Athanassoula, \emph{Mon. Not. R. Astron. Soc.}, \textbf{259}, 354, (1992b).

%\bibitem{ZS1}A. El-Zant and I. Shlosman, \emph{ApJ}, \textbf{577}, 626, (2002).
%\bibitem{ZS2}A. El-Zant and I. Shlosman, \emph{ApJ}, \textbf{595}, L41, (2003).









%\bibitem{Brown2000} M.~P. Brown,  and K.~Austin, \emph{The New Physique}, Publisher Name, Publisher City, 2000, pp. 212--213.

%\bibitem{BrownAustin:2000} M.~P. Brown,  and K.~Austin, \emph{Appl. Phys. Letters}, \textbf{85},  2503--2504 (2000).

%\bibitem{Wang} R.~Wang, ``Title of Chapter,'' in \emph{Classic Physiques}, edited by R.~B. Hamil, Publisher Name, Publisher City, 2000, pp. 212--213.

%\bibitem{SJ:1999} C.~D.~Smith and E.~F.~Jones,  ``Load-Cycling in Cubic Press,'' in \emph{Shock Compression of Condensed Matter-1999}, edited by M.~D.~F. et~al.,  AIP Conference Proceedings 505, American Institute of Physics, New York, 1999, pp. 651--654.

\end{thebibliography}

%%%%%%%%%%%%%%%%%%%%%%%%%%%%%%%%%%%%%%%%%%%
%% Just a reminder that you may have to run bibtex
%% All of it up to \end{document} can be removed
%% if you don't like the warning.
%%%%%%%%%%%%%%%%%%%%%%%%%%%%%%%%%%%%%%%%%%%
\IfFileExists{\jobname.bbl}{}
 {\typeout{}
  \typeout{******************************************}
  \typeout{** Please run "bibtex \jobname" to optain}
  \typeout{** the bibliography and then re-run LaTeX}
  \typeout{** twice to fix the references!}
  \typeout{******************************************}
  \typeout{}
 }

\end{document}